\documentclass[preprint,showpacs,preprintnumbers,amsmath,amssymb]{revtex4}
\usepackage{graphicx}% Include figure files
\usepackage{dcolumn}% Align table columns on decimal point
\usepackage{bm}% bold math
\usepackage{longtable}
\begin{document}

\title{Ferromagnetism in Mn doped GaAs due to substitutional-interstitial complexes}
\author{Priya Mahadevan and Alex Zunger \\
National Renewable Energy Laboratory, Golden-80401}
\date{\today}

\begin{abstract}
While most calculations on the properties of the ferromagnetic semiconductor GaAs:Mn have focussed 
on isolated Mn substituting the Ga site (Mn$_{Ga}$), we investigate here whether alternate lattice
sites are favored and what the magnetic consequences of this might be.
Under As-rich (Ga-poor) conditions prevalent at growth, we find that 
the formation energies are lower for Mn$_{Ga}$ over interstitial Mn (Mn$_i$).
As the Fermi energy is shifted towards the valence band maximum via external $p$-doping, the 
formation energy of Mn$_i$ is reduced relative to Mn$_{Ga}$. Furthermore,
under epitaxial growth conditions, the solubility of both substitutional and interstitial Mn are strongly 
enhanced over what is possible under bulk growth conditions. The high concentration of Mn attained under
epitaxial growth of $p$-type material opens the possibility of Mn atoms forming small clusters. We consider 
various types of clusters, including the Coulomb-stabilized clusters involving two Mn$_{Ga}$ and one Mn$_i$.
While isolated Mn$_i$ are hole killers (donors), and therefore destroy ferromagnetism, 
complexes such as (Mn$_{Ga}$-Mn$_i$-Mn$_{Ga}$) are found to be 
more stable than complexes involving Mn$_{Ga}$-Mn$_{Ga}$-Mn$_{Ga}$. The former complexes exhibit partial or total
quenching of holes, yet Mn$_i$ in these complexes 
provide a channel for a ferromagnetic arrangement of the spins on the two Mn$_{Ga}$ within the complex. 
This suggests that ferromagnetism in Mn doped GaAs arises both from holes due to isolated Mn$_{Ga}$
as well as from strongly Coulomb stabilized Mn$_{Ga}$-Mn$_i$-Mn$_{Ga}$ clusters. 
\end{abstract}

%]

%\narrowtext
\maketitle
\newpage

\section {Introduction}

The discussion \cite{theory,cluster} of the physics that underlies
room-temperature ferromagnetism in transition-metal doped
semiconductors has largely focussed on {\it substitutional} 
geometries, e.g. Mn$_{Ga}$ site in GaAs. Indeed, there is a
well-established tradition that 3$d$ impurities in III-V
semiconductors are largely substitutional \cite{subst}, while in Si they are mostly
interstitials \cite{Ludwig}. Modern first-principles total-energy
calculations afford testing of this classic paradigm. Recent experiments \cite{rbs} find that 
Mn atoms occupy both substitutional as well as interstitial positions in GaAs. 
There have been suggestions from 
recent theoretical work \cite{erwin} suggested that primarily 
surface energetics will funnel Mn atoms in to 
interstitial sites from surface adatom positions. 
While Mn$_{Ga}$ behaves as a hole-producing acceptor, at the interstitial site, Mn$_i$ behaves as an
electron-producing donor. Since ferromagnetism is mediated by free-carriers, Mn$_i$ could 
modify the magnetic properties from the case where only substitutional Mn sites were occupied. 

Using density-functional 
theory as implemented within plane-wave pseudopotential total energy method, we consider here
bulk and epitaxial growth conditions, investigating isolated defects (Mn$_{Ga}$ and Mn$_i$) and their complexes.
We find that the Mn impurity in GaAs is stable in both substitutional and interstitial 
geometries depending on (a) the Fermi energy (which can be changed via external doping), (b) chemical potentials during growth 
and (c) bulk versus epitaxial growth conditions. 
The origin of these dependences is as follows: (a) As the formation energy of impurities 
that are neutral with respect to the lattice site they occupy ({\it e.g.} Mn$^0_{Ga}$) does not
depend on the Fermi energy ($\epsilon_F$), the formation energy of positively 
charged impurities ({\it e.g.} Mn$_i^{2+}$)
decreases as $\epsilon_F$ is shifted towards the VBM. Hence, the difference in the formation 
energies between Mn$_i^{2+}$ and Mn$_{Ga}^0$ decreases with $p$-type doping, resulting in increased solubility 
of interstitial Mn.
(b) Substitution of Ga by Mn involves the removal of a Ga atom and the introduction of a Mn atom at 
the site vacated by Ga. Thus, substitution is generally enhanced under Ga-poor, Mn-rich growth conditions.
On the other hand the formation energy of Mn at
an interstitial site does not depend on the Ga chemical potential. Thus, one may stabilize substitutional 
(interstitial) doping using Ga-poor (Ga-rich) growth conditions.
(c) Solid solubility can be controlled thermodynamically using epitaxial instead of bulk
growth conditions \cite{wood}. The absence of a substrate under bulk growth conditions
allows the growing solid as well as its possible disproportionation products 
to attain their free-standing lattice geometry.
This is the case when the growth takes place from the melt as in Bridgman growth.
Then, if phase separation occurs, the precipitate will take up
its most stable crystal structure, {\it i.e.} MnAs in the NiAs structure.
In contrast, under thin-film epitaxial growth conditions (as in MBE, MOCVD) competing phases such as phase 
separated MnAs are forced to be coherent with the GaAs substrate.
As zincblende MnAs strained on GaAs is less stable than the NiAs phase of MnAs, phase separation is more
costly under coherent epitaxial conditions, and one expects \cite{wood}
less phase separation, hence enhanced solubility. We find the following:

(i) Substitutional Mn has two stable charge states: neutral (Mn$_{Ga}^0$) and negatively charged
(Mn$_{Ga}^-$) charge state which have 1 and 0 holes, respectively. 
The calculated acceptor transition E(0/-) between these states occurs at E$_v$ +0.13 eV, in good
agreement with the experimental value of E$_v$+0.11 eV \cite{schneider}. Here E$_v$ corresponds to 
the valence band maximum of the host material.

(ii) The interstitial sites that Mn can occupy have either tetrahedral (coordinated to four As or four Ga atoms) or 
hexagonal symmetry.
We find that Mn at the tetrahedral interstitial site coordinated by As is more
stable than that coordinated by Ga, and exhibits a single charge state Mn$^{2+}$ for all values of the 
Fermi energy. The (0/+) and (+/2+) donor transitions are found to lie inside the conduction band, so,
isolated Mn$_i$ produce electrons that will compensate the holes created by Mn$_{Ga}$. 

(iii) Under {\it bulk} growth conditions, the formation energy per Mn 
of substitutional Mn is $\Delta$H(Mn$^0_{Ga}$)=0.91+$\mu_{Ga}$-$\mu_{Mn}$ eV=0.17~-$\mu_{As}$-$\mu_{Mn}$ eV,
whereas interstitial Mn has $\Delta$H (Mn$^{2+}_i$)=0.55-$\mu_{Mn}$+2$\epsilon_F$ eV per Mn. Here
$\epsilon_F$ is the Fermi energy measured with respect to the valence band maximum of the host material, 
and $\mu_{As}$ and $\mu_{Mn}$ are the chemical potentials of As and Mn respectively.
If we use maximally As-rich growth conditions ($\mu_{As}$=0 eV) and 
$\mu_{Mn}$=$\Delta$H(MnAs), then we find $\Delta$H(Mn$_{Ga}^0$)=0.91 eV.
As one dopes the sample $p$-type and $\epsilon_F$
approaches the VBM ($\epsilon_F$=0), the energy difference between the formation energies 
of Mn$_{Ga}^0$ and Mn$_i^{2+}$ reduces to 0.38 eV. It could decrease even further if
$\epsilon_F$ penetrates the valence band with doping, 
or if the growth conditions are made less As-rich. The interstitial 
concentration is then expected to further increase. This is confirmed by recent 
experiments \cite{lpe} which use Ga-rich conditions for growth.

(iv) Under {\it epitaxial} growth conditions, the formation energies of both substitutional and interstitial 
Mn decrease by 0.74~eV/Mn, so their concentrations increase concomitantly leading to the possibilities 
of clusters. There is strong coulomb interactions between the oppositely charged constituents 
involving two substitutional (Mn$_{Ga}$) and one interstitial (Mn$_i$); the cluster 
Mn$_{Ga}$-Mn$_i$-Mn$_{Ga}$ is thus strongly stabilized and found
to be more stable under $p$-type conditions than clusters involving three Mn$_{Ga}$.
Epitaxial growth conditions increases the solubility of such Mn$_{Ga}$-Mn$_i$-Mn$_{Ga}$ clusters, with 
formation energy of -0.15 +$\epsilon_F$ eV per cluster for the Q=+1 charge state under As-rich conditions and
$\mu_{Mn}$=$\Delta$H(MnAs). 

(v) The presence of interstitial Mn in 
the Mn$_{Ga}$-Mn$_i$-Mn$_{Ga}$ cluster provides a channel for the spins on the two 
substitutional Mn to align ferromagnetically even when 
there are no free carriers present in the cluster. We therefore conclude that ferromagnetism in GaAs:Mn
can arise both from holes induced by isolated substitutional Mn atoms  discussed previously (Dietl) as well as from 
charge compensated substitutional-interstitial clusters.

\section{Method of Calculation}

The formation energy for a defect comprising of atoms $\alpha$ in the charge
state $q$ was computed using the 
density functional supercell method using the expression \cite{cuinse2}
\begin{eqnarray}
{\Delta}H_f^{\alpha,q}(\epsilon_F,\mu)& =& E(\alpha) - E(0) +
\sum_{\alpha} n_{\alpha} \mu_{\alpha}^{a} + q (E_{v} +
\epsilon_F),
\end{eqnarray}
where  $E(\alpha)$ and $E(0)$ are the total energies
of a supercell with and without the defect $\alpha$ respectively. 
$n_{\alpha}$ denotes the number of atoms of defect $\alpha$
transferred in or out of the reservoir (equal to 1 for an atom removed and -1
for an atom added), while $\mu_{\alpha}^a$ denotes their chemical potentials.

{\it Total energies:} The total energies of the
charged supercells were computed by compensating any additional charge on the
impurity atom by a uniform jellium background and 
have been corrected for interactions between charges in neighboring cells 
using the Makov and Payne correction \cite{payne}.
For isolated defects we used both the monopole as well as
quadrupole corrections, while for composite defects we have added only the monopole correction to the total energy
assuming all the charge to be localized at a single point. 
We use the static dielectric constant of GaAs (12.4) \cite{madelung}.
The quadrupole moment of the isolated defects was calculated as the difference between the moments of the
supercell with the charged defect and that with the neutral defect.

{\it Transition energies:} The defect transition energy $\epsilon(q,q')$ is the value of the Fermi
energy $\epsilon_F$ at which ${\Delta}H^{\alpha,q}(\epsilon_F)$=${\Delta}H^{\alpha,q'}(\epsilon_F)$. The
zero of the Fermi energy is chosen as the valence band maximum
$E_{v}$ of the pure host at the $\Gamma$ point.

{\it Chemical potential limits:} As the reservoir supplying
the atoms could be elemental solids, or compounds formed from the
elements, we express $\mu_{\alpha}^a$ as the sum of the energy of
the element in its most stable structure $\mu_{\alpha}^s$, and an
additional energy $\mu_{\alpha}$ {\it i.e} $ \mu_{\alpha}^a$ =
$\mu_{\alpha}^s$ + $\mu_{\alpha}$. The required ranges of
$\mu_{\alpha}$ are determined by $\mu_{Ga}$ $\le$ 0; $\mu_{Mn}$
$\le$ 0; $\mu_{As}$ $\le$ 0 (no precipitation of solid elements)
and by the formation energies of GaAs, and MnAs. 
The allowed values of chemical potential are such that GaAs is stable, {\it
i.e.} $\mu_{Ga}+\mu_{As}={\Delta}H_f(\text{GaAs})$, the latter being the
formation energy of zinc-blende GaAs. Further, as Mn should not
precipitate  as MnAs, we restrict $\mu_{Mn}+\mu_{As}<{\Delta}H_f(\text{MnAs})$, 
the formation energy of MnAs in its
most stable (NiAs) structure. For epitaxial growth conditions, the formation 
energy of zinc-blende MnAs lattice-matched to GaAs is considered. In this case we calculate the epitaxial
formation energy, ${\Delta}H_f(\text{MnAs})_{epi}$, forcing the in-plane lattice constant of MnAs to
equal that of GaAs, while the out-of-plane lattice constant, c,  is allowed to vary.
For coherent epitaxial growth the condition that MnAs should not form during incorporation of Mn in GaAs
becomes $\mu_{Mn}+\mu_{As}<{\Delta}H_f(\text{MnAs})_{epi}$.

The energies E($\alpha$), E(0), ${\Delta}H_f(\text{GaAs})$, ${\Delta}H_f(\text{MnAs})$, 
${\Delta}H_f(\text{MnAs})_{epi}$ and $\mu_{\alpha}$ are calculated
within the density functional formalism, through the momentum-space pseudopotential total energy 
representation \cite{ihm}, using ultrasoft pseudopotentials \cite{ultrasoft}. 
The GGA-PW91 version of the exchange-correlation functional \cite{pw91} and no 
correction for the band gap underestimation was made.
The calculations were performed over a Monkhorst-Pack 4x4x4 k-point grid for 
64 \cite{compare} and 216 atom supercells of GaAs using VASP \cite{vasp}. Changing the k-point mesh
from 2x2x2 to 4x4x4 changed the formation energies by $\sim$ 20~meV. Larger 256 atom supercells with 1x1x2 k-points were 
used for the calculations with clusters to ensure a larger separation between clusters.
We used a plane wave cutoff of 227.2 eV for these calculations. Increasing the cutoff to 300 eV,
changed the formation energies by $\sim$ 10 meV.
As the lattice constant of the supercell was kept fixed at the GGA optimised value for GaAs of $a$=5.738 $\AA$ \cite{footnote1},
the internal coordinates were optimised. Our calculated (experimental) formation energies are
${\Delta}H_f(GaAs)$~=~-0.74 (-0.74), ${\Delta}H_f(MnAs)$~=~-0.74 (-0.61) eV and  
${\Delta}H_f(MnAs)_{epi}$~$\sim$ 0 eV.
For elemental Mn we assume the nonmagnetic
fcc structure \cite{mn_reference}, 
while for elemental Ga, we assume the base-centered orthorhombic structure.

The charge corrected \cite{payne} Mn$_{Ga}$ (0/-) transition as well as the difference in formation 
energies between Mn$_{Ga}^0$ and Mn$_i^{2+}$ are
given in Table I for supercell sizes of 64 and 216 atoms. We see that changing the supercell size from 64 to 216 atoms
lowers the acceptor energy by 30-50 meV and stabilizes Mn$_{Ga}^0$ over Mn$_i^{2+}$ by 50-150 meV. The
charge correction increases the acceptor energy by 60-90 meV and stabilizes Mn$_{Ga}^0$ over Mn$_i^{2+}$ by 
250-350 meV.

\section{Results} 

{\subsection{Isolated substitutional Mn on the Ga site of GaAs}}

Fig. 1 describes the formation energy $\Delta$H(Mn$_{Ga}^0$) of neutral substitutional Mn in GaAs as a 
function of the chemical potentials $\mu_{As}$ and $\mu_{Mn}$.
The shaded areas denote chemical potentials that produce unwanted products: (i) When $\mu_{As}$ becomes greater than
zero (the cohesive energy of solid As) we have precipitation of elemental As as shown 
on the left hand side of Fig.~1. (ii) In the opposite limit, when $\mu_{As}$ takes more negative values than the formation 
energy $\Delta$H(GaAs), we have maximally As-poor conditions and the host itself becomes unstable, as
shown on the right hand side of Fig.~1. (iii) The diagonal lines in the main body of Fig.~1 denote different values 
of $\mu_{Mn}$. When the chemical potential of Mn becomes greater than zero (the cohesive energy of solid Mn), metallic Mn
will precipitate as shown in the bottom right corner of Fig.~1. Conversely, (iv) when $\mu_{Mn}$ becomes
equal or larger than $\Delta$H(MnAs)-$\mu_{As}$, we will precipitate a secondary phase of MnAs. Clearly, since 
$\mu_{Ga}$+$\mu_{As}$=$\Delta$H(GaAs) and $\mu_{Mn}$+$\mu_{As} <$~$\Delta$H(MnAs), one can keep the latter 
inequality even for moderately negative values of $\mu_{Mn}$, provided that $\mu_{As}$ is adjusted.
The lines in Fig.~1 show that the lowest $\Delta$H(Mn$_{Ga}^0$) value is 0.91~eV (circle at bottom
left corner). This can be attained 
at $\mu_{As}$=0 (maximally As-rich); $\mu_{Mn}$=$\Delta$H(MnAs). Alternatively, the same 
solubility can be attained for less rich-As conditions, but richer
Mn conditions, {\it e.g.} for $\mu_{As}$=-0.5 eV and $\mu_{Mn}$=-0.24 eV.

Having described in Fig.~1 the stability of the {\it neutral} substitutional, we 
next describe in Fig.~2 the stability of the
{\it charged} substitutionals. Here we chose the chemical potentials $\mu_{As}$=0, $\mu_{Mn}$=$\Delta$H(MnAs) (denoted by the
circle in Fig.~1) and vary the Fermi energy. We see that for $p$-type conditions, the lowest energy charge state is Mn$_{Ga}^0$,
whereas for higher Fermi energy the stablest charge state is Mn$_{Ga}^-$. Table I gives the (0/-) acceptor transition energy
calculated with various supercell sizes with and without charge correction. The most converged (0/-) transition energy
calculated for the 216 atom cell and corrected for charge interactions is E$_v$+0.13~eV, in good agreement with the
measured value of E$_v$+0.11~eV \cite{schneider}. Fig.~2 shows that under epitaxial conditions (right y-axis), the
formation energy of Mn$_{Ga}^0$ is lowered by 0.74~eV.

We next describe the electronic structure of Mn$_{Ga}$.
In Figs. 3(a) and (b) we show the Mn $d$ projected partial density of states (PDOS) for two 
charge states of substitutional Mn. The main features can be understood as 
arising from the hybridization between 
the anion dangling bonds generated by a Ga vacancy and the $d$ levels on the Mn ion placed at the vacant 
site \cite{subst}. The Mn $d$ ion levels are split by the tetrahedral crystal field into
$t_2(d)$ and $e(d)$. Exchange interactions further split these levels into spin-up ($\uparrow$)
and spin-down ($\downarrow$) levels. The $t_2(d)$ levels on the Mn atom hybridize with the levels with the same 
symmetry on the As dangling bonds, while the $e(d)$ levels have no other states available for significant 
coupling \cite{subst}. Because  the location of the Mn ion $d$ levels is below the dangling bond levels, 
after hybridization, the deeper bonding $t_2$ states have dominantly Mn $d$ character (refered to as CFR: 
"crystal field resonance"),
while the higher antibonding $t_2$ states have dominantly As $p$ character (refered to as DBH : "dangling bond hybrid").
These interactions lead to the energy level diagram depicted schematically on the left-hand-side of
Fig.~4 showing a fully 
occupied, Mn-localized up-spin CFR of $t_2$ and $e$ symmetries.
At  a higher energy we have the  up and down-spin 
DBH states with $t_2$ symmetry. Because of the location of the Ga vacancy states $t_2(p)$
between the exchange split $t_2(d)$ states on the Mn, a negative 
exchange splitting is induced as a result of hybridization on the DBH states \cite{our_paper} with 
$t^{\downarrow}_{DBH}$ below $t_{DBH}^{\uparrow}$. As a result, the neutral substitutional
defect Mn$^0_{Ga}$ has the electron configuration
[$t_{\uparrow}^3e_{\uparrow}^2$]$_{CFR}$ ($t_{\downarrow}^3t_{\uparrow}^2$)$_{DBH}$, with a total
magnetic moment $\mu$=4 $\mu_B$, and a hole in the 
$t_{DBH}^{\uparrow}$ orbital. This configuration corresponds to
the multiplet $^5$T$_2$ as observed in electron paramagnetic resonance (EPR) experiments \cite{mn_acc}. 
The partial occupancy of the negative
exchange-split DBH states stabilizes the ferromagnetic state over
the antiferromagnetic state \cite{our_paper}. 

{\subsection {Isolated interstitial Mn}} 

Mn interstitial can occupy a site with tetrahedral symmetry (coordinated by four As or four Ga atoms)
or a site with hexagonal symmetry. We have calculated the total energies of Mn at these positions in a 64 atom cell of GaAs, 
and the results for the tetrahedral interstitial sites are given in Table II.
The tetrahedral interstitial Mn$_i$(As) coordinated by four As atoms is more stable
than the one coordinated by four Ga atoms, with the difference being 
0.16, 0.31 and 0.31 eV for charge states $q$=1,2 and 3.
In contrast, the hexagonal interstitial has 0.62 eV higher total 
energy than the most stable  Mn$^{2+}_i$(As). Experimentally, the presence 
of interstitial Mn was detected by an analysis of the EPR spectrum 
\cite{epr} as well as by Rutherford back scattering \cite{rbs}.
The distinction between the two types of $T_d$ interstitial sites
(Mn-next to As vs Mn-next to Ga) is difficult to determine experimentally and
involved an analysis of the experimentally measured 
contact interaction in
terms of the covalency of the Mn-X bond. This analysis suggested that Mn$_i$(Ga)
was more stable, while our total energy calculations suggest that Mn$_i$(As) is 
more stable. 

The formation energy of various charge states of interstitial Mn is shown in Fig.~2
for $\mu_{As}$=0 and $\mu_{Mn}$=$\Delta$H(MnAs). We see that the stable charge state is Mn$_i^{2+}$
for the full range of Fermi level, with maximum stability at $\epsilon_F$=0. To compare the relative 
stability of Mn$_i^{2+}$ at $\epsilon_F$=0 with substitutional Mn$_{Ga}^0$, we show in the upper scale
of Fig.~1 the difference $\Delta$H(Mn$_i^{2+}$) - $\Delta$H(Mn$_{Ga}^0$) between the formation
energies of interstitial and substitutional Mn. We see that substitutional Ga is
stabler on the left hand side of the figure, {\it i.e} sufficiently As-rich, whereas 
interstitial Mn is stabler at the right hand side of the figure, {\it i.e.} sufficiently As-poor.
The energy difference is 
$${\Delta}H(Mn_i^{2+}) - {\Delta}H(Mn_{Ga}^0)=0.38 + \mu_{As} + 2\epsilon_F$$
For $\mu_{As}$=0, the substitutional Mn are stabler by 0.38~eV, while for moderately As-rich conditions, say 
$\mu_{As}$ = -0.4~eV, both defects have comparable formation energies.

These results are in agreement with recent experiments using liquid phase epitaxy \cite{lpe} to
introduce Mn in GaAs. Experimentally a decrease in hole concentration is found as the
Mn concentration is increased. Under the Ga-rich growth conditions used, As antisites are not expected
to be the dominant source of the observed compensation. 
Hence the major source of compensation is believed to come from
Mn$_i$ as expected for Ga-rich conditions from Fig. ~1.

We next examine the electronic structure of Mn at a
tetrahedral interstitial site. When Mn occupies a tetrahedral interstitial position, 
five of the seven electrons occupy the $t_{\uparrow}e_{\uparrow}$ CFR levels, with the remaining two
going into the down-spin $t_{\downarrow}$ levels. This is evident from the PDOS for the doubly-ionized
Mn$_i^{2+}$ shown in Fig. 3(c), 
where Mn$_i^{2+}$ is found to have the configuration [$t_{\uparrow}^3e_{\uparrow}^2$]$_{CFR}$ with
a magnetic moment of $\mu$=5$\mu_B$. The central panel of Fig. 4(a) shows schematically the 
levels of Mn$_i$. As the (0/+) and (+/2+) transitions are calculated to lie inside the GaAs
conduction band (Fig. 2), we conclude that Mn$_i$ produce free electrons in GaAs. 

{\subsection {Clusters of substitutional Mn}} 

Having dealt with the
isolated limit, we investigated whether Mn atoms show a tendency
to cluster. Recent experiments \cite{cocluster} on dilute magnetic semiconductors have found a strong tendency 
of the doped transition metal atoms to cluster and there has been some theoretical work \cite{cluster} 
to support such observations. 
We consider As-centred clusters [(As)Mn$_n$Ga$_{4-n}$] with $n$=0,1,2,3 and 4.

Fig. 5 shows the formation energy of clusters made of three substitutional Mn atoms (S-S-S) at 
lattice locations (0,0,0), (a/2,a/2,0) and (0,a/2,a/2) in the 64 atom supercell. This corresponds to the 
$n$=3 cluster. Here $a$ is the 
cubic lattice constant of GaAs. We see that the neutral cluster (3~Mn$_{Ga}$)$^0$ having three holes is stable 
under $p$-type conditions, whereas the charged cluster (3Mn$_{Ga}$)$^-$ with 2 holes is more stable above 
$\epsilon_F$ = 0.15 eV. 
The energies of the complex with 3(Mn$_{Ga}$)$^0$ is 2.2 + 3($\mu_{Ga}$ - $\mu_{Mn}$) eV
while that of three noninteracting Mn$_{Ga}$ in their lowest energy charge state is 
2.71 + 3($\mu_{Ga}$ - $\mu_{Mn}$) eV.
For epitaxial conditions $\Delta$H$_{epi}$=0.02 +3($\mu_{Ga}$ - $\mu_{Mn}$) eV/cluster. Thus, as
the formation energy is very low, 
the tendency for the Mn atoms to cluster is strongly enhanced under epitaxial growth conditions.

In order to obtain a measure of the tendency to cluster, we calculate the clustering energy.
The "clustering energy" $\delta(n)$ is defined as the energy
difference between n substitutional Mn atoms surrounding an As
site [(As)Mn$_n$Ga$_{4-n}$; $0 \le n \le 4$] and n isolated well-separated
constituents.
Thus, $\delta E(n)=[E(n)-E(0)]-n[E(1)-E(0)]$, where E(n) is the
total energy of the supercell with As-centered clusters of n Mn 
atoms. We find that $\delta E$(n)= -228, -482 and -794
meV per cluster of n=2,3 and 4 Mn atoms for a 64 atom supercell.
The clustering energy changed to -519, -1069 meV for clusters involving 
2 and 3 Mn in a 256 atom supercell.
These results indicate a strong tendency for the neutral substitutional Mn atoms 
to form clusters. 

{\subsection {Neutral complexes of Mn$_{Ga}$ and Mn$_i$}} 

We considered 
the defect complex formed between Mn$_{Ga}$ (S) and 
interstitial Mn$_i$ (I) denoted as (S-I-S)$^Q$, where $Q$ is the total charge of the complex. 
The geometry for the complex had Mn$_{Ga}$ at (0,0,0) and (a/2,a/2,0) and Mn$_i$ at (a/2,0,0) in 
the 64 atom supercell of GaAs.
We see in Fig. 5 that S-I-S exists in two charge states: When the Fermi energy is below E$_v$+0.1 eV 
we have the stable structure is (S-I-S)$^{1+}$, whereas when $\epsilon_F$ is above it, the stable 
structure is the neutral (S-I-S)$^0$. Thus, the donor transition for the cluster is at 
E$_v$+0.1 eV. Fig. 5 also shows that for Fermi levels 
below E$_v$+0.22 eV, the S-I-S complex is stabler than the S-S-S 
complex. As for the interaction energy of the components of the complex:
the formation energy of the non-interacting neutral ($Q=0$) components of the complex is
2E(Mn$^0_{Ga}$)+E(Mn$_i^0$)=4.31+2$\mu_{Ga}$~-~3$\mu_{Mn}$ eV per 3 impurities,
while the formation energy of the {\it
interacting} neutral complex is 1.41+2$\mu_{Ga}$-3$\mu_{Mn}$ eV. 
This represents a $\sim$ 2.9 eV per 3 impurities stabilization over the
non-interacting, neutral defects. The energy of the neutral complex measured with
respect to the stablest lattice site occupied by isolated Mn under particular experimental conditions 
is found to be -574 meV for $\mu_{As}$=0 eV, $\mu_{Mn}$=$\Delta$H(MnAs) and $\epsilon_F$=0 eV.
Hence this complex is strongly stabilized.

The reasons for the stability of the (S-I-S)$^0$
complex can be appreciated from Fig. 4(a). 
Upon bringing together 2Mn$_{Ga}^0$ with Mn$_i^0$, one 
electron drops from the higher energy $t_{\downarrow}^{CFR}$ level of 
Mn$_i$ to the lower energy $t_{\uparrow}^{DBH}$ level of each
substitutional site, resulting in 
$[t_{\uparrow}^3e_{\uparrow}^2]_{CFR}(t_{\downarrow}^3t_{\uparrow}^3)_{DBH}$ configuration at each
Mn$_{Ga}$ site (Fig. 4(b)) which corresponds to Mn$_{Ga}^-$. These conclusions
are evident from our calculated DOS of the S-I-S complex, projected 
on the I and S sites shown in Fig. 6. We find that for both $Q$=0 (Fig. 6(a)) 
and $Q$=2 (Fig. 6(b)) the I site has the configuration 
$[t^3_{\uparrow} e_{\uparrow}^2]_{CFR}$
or "$d_{\uparrow}^5$".
This substitutional-to-interstitial charge transfer lowers the energy
of the complex by twice the separation between $t_{\downarrow}^{CFR}$ level
of Mn$_i$ and $t_{\uparrow}^{DBH}$ level of Mn$_{Ga}$. Furthermore, it
creates a favorable Coulomb attraction between the components
S$^-$-I$^{2+}$-S$^-$ of the complex. 
This energetically favorable
substitutional-interstitial association reaction then eliminates
the holes that were present in isolated substitutional Mn$_{Ga}$
and could explain the puzzling observation
\cite{hole_conc} of the existence of a far lower concentration of
holes than Mn in GaAs.
Alternate explanations such as the
presence of As antisites \cite{hole_mech} as well as the presence of Mn atoms connected to six
As atoms (as in the NiAs structure) have been
offered. However, samples have been prepared where
the concentration of As antisites is too low to explain
the observed compensation of holes.
Further, six-fold coordinated Mn atoms have not been observed in Mn
doped GaAs samples \cite{exafs}.

{\subsection {Ferromagnetism of the $(S-I-S)^0$ complex} }

The neutral complex has two Mn$_{Ga}^-$  and one intervening "$d_{\uparrow}^5$" interstitial. 
We find that a ferromagnetic arrangement between Mn$_{Ga}$ is favored in the 
complex ($Mn_{Ga}^{-}-Mn_i^{2+}-Mn_{Ga}^{-}$)$^0$ by 176 meV.  In contrast, our calculations for two Mn$_{Ga}^-$ atoms
{\it without} the intervening interstitial atom finds that an {\it antiferromagnetic} arrangement of spins on the
substitutional Mn atoms is favored by 108 meV. Thus Mn$_i$ is responsible for mediating 
a ferromagnetic interaction between Mn$_{Ga}^0$.

How does the presence of the interstitial Mn mediate the alignment of spins on the
substitutional Mn? There are three possible arrangements for the spins on the Mn atoms making up the neutral complex
- (S$^{\uparrow}$I$^{\uparrow}$S$^{\uparrow}$)$^0$, 
(S$^{\uparrow}$I$^{\downarrow}$S$^{\uparrow}$)$^0$ and 
(S$^{\uparrow}$I$^{\downarrow}$S$^{\downarrow}$)$^0$.
From our total energy calculations we find that the energy 
for the configurations (S$^{\uparrow}$I$^{\uparrow}$S$^{\uparrow}$)$^0$ and  
(S$^{\uparrow}$I$^{\downarrow}$S$^{\downarrow}$)$^0$ 
are higher by 563 meV and 176 meV, respectively, than the energy, E$_0$,
of the ground state (S$^{\uparrow}$I$^{\downarrow}$S$^{\uparrow}$)$^0$. 
(The energies changed marginally to 602~meV and 192~meV respectively when we 
increased the supercell size to 256 atoms.)
The stabilization of the S$^{\uparrow}$I$^{\downarrow}$S$^{\uparrow}$ magnetic arrangement can be understood
using simple arguments:
In the configuration (S$^{\uparrow}$I$^{\uparrow}$S$^{\uparrow}$)$^0$, 
as one spin channel is completely filled, there is no channel of hopping 
available for the electrons to delocalize and lower their energy. Thus, this is a
high energy spin configuration with energy E$_0$+563 meV. In contrast, 
in the configuration (S$^{\uparrow}$I$^{\downarrow}$S$^{\downarrow}$)$^0$,
two channels of hopping are present; the first between the electrons on S$^{\uparrow}$ and 
S$^{\downarrow}$, and the second between those on
S$^{\uparrow}$ and I$^{\downarrow}$. This configuration is found to have the energy, E$_0$+176 meV.
Likewise, the configuration (S$^{\uparrow}$I$^{\downarrow}$S$^{\uparrow}$) which has energy E$_0$ has
two channels of hopping present between S and I.
The dominant factor in determining the configuration which
has the lowest energy are the hopping matrix elements - V$_{S,I}$ between S and I and V$_{S,S}$ between the two S.
To a first approximation, these hopping matrix elements are determined by the separation between the atoms
involved.
As the distance between the two substitutional Mn atoms is $\sqrt{2}$ times the distance between
S and I, the effective hopping matrix element between S is smaller.
Hence the presence of
an intervening Mn$_i$ provides a channel for the  ferromagnetic arrangement of spins between two Mn$_{Ga}$ even in the
neutral charge-compensated complex. 
In contrast, the presence of a closed shell donor such as As$_{Ga}$ between 
two Mn$_{Ga}$ gives rise to an 
antiferromagnetic (or weakly ferromagnetic) interaction between Mn$_{Ga}$.  

How does the presence of the interstitial affect long-range ferromagnetism? In order to investigate this
we introduced a hole-producing, isolated substitutional Mn atom at different lattice 
locations, indicated in Fig. 7, and investigated whether the spin on this isolated 
substitutional Mn atom prefers to align parallel or antiparallel with respect to the spins on the substitutional Mn 
atoms within the S-I-S cluster.
We find that the substitutional Mn likes to align ferromagnetically with the Mn$_{Ga}$ of the cluster by 147, 214 and 81 meV,
respectively for the positions 1, 2 and 3 
(see Fig. 7). Hence the presence of the interstitial Mn forces a hole which is 
located $\sim$ 12 $\AA$ from the S-I-S cluster to align ferromagnetically and therefore contributes 
to the long-ranged ferromagnetism observed in these systems.

As discussed earlier, the basic electronic structure of substitutional Mn in GaAs can be understood as arising
from the hopping interaction between the Mn $d$ states and the As $p$ dangling bond states. Therefore, 
the coupling between  two Mn atoms is through the As $p$ states. It is strongest 
along the directions in which the "$p-d$" coupling of the Mn with the  As states is the
largest and decreases with distance along that direction. 
When there is a hole, the antibonding $t_2^{\uparrow}$ orbitals are partially occupied, and there is ferromagnetism.
Hence in Fig. 7 spins on the Mn atoms at
sites 2 and 3 prefer to align ferromagnetically with the spins on S when there is partial compensation. 
This could happen either because the S-I-S cluster is totally compensated, and a hole exists on the Mn at sites 2/3 or
the S-I-S cluster is partially compensated.
The mechanism stabilizing the ferromagnetic coupling between S and site 1 is the same as what we
discussed for the {(S-I-S)$^0$} complex earlier and exists even when there is total compensation.

{\subsection {Charged Mn$_{Ga}$-Mn$_i$-Mn$_{Ga}$ complexes:}} 

While the neutral complex has no holes, the
$Q$=+1 and +2 complexes have 1 and 2 holes respectively (Fig. 4(c)) 
and a net magnetic moment of 4 and 3 $\mu_B$
respectively.  For $Q$=+2, Mn$_{Ga}$ adopts the configuration Mn$_{Ga}^0$ (Fig. 4(c)).
We find that (Mn$_{Ga}$-Mn$_i$-Mn$_{Ga}$)$^{2+}$ prefers the ferromagnetic arrangement of spins on Mn$_{Ga}$
by 286~meV, similar to the ferromagnetic preference ($\sim$ 305 meV) 
of Mn$_{Ga}^0$-Mn$_{Ga}^0$ pair without an intervening Mn$_i$. These results suggest the surprising fact 
that the spins on Mn$_{Ga}$ align ferromagnetically in the charged complexes, almost as if Mn$_i$ did not exist.
The number of holes in the cluster is the same as the number in the pair, though the number of Mn atoms
are different. This is in agreement with the experimental observation \cite{mag_deficit} where above a critical 
concentration of Mn, both the number of holes as well as the ferromagnetic transition temperature remain constant , while
the magnetic moment per Mn atom decreases \cite{mag_deficit}.
The magnetic moments that we obtain for the $Q$=+1 and +2 charge states translate into average moments of 1.33 $\mu_B$
and 1 $\mu_B$ per Mn, while the uncompensated pair of Mn$_{Ga}$ have a magnetic moment of 4 $\mu_B$. 
In this regime where the $T_c$ is found to saturate, the average magnetic moment per Mn is found to 
vary from $\sim$ 3 $\mu_B$ at a Mn concentration of 5.54 $\%$ to 1.74 at 8.3 $\%$. 

Recent experiments \cite{waluck2} find that the $T_c$ of the as-grown samples 
increased after annealing. This was interpreted as the migration of FM-reducing {\it interstitial} Mn to FM-enhancing {\it substitutional}
positions. We investigated which clusters could break by annealing and promote ferromagnetism. 
As the S-I-S complexes are rather strongly bound with respect to their
constituents, we investigated instead complexes S-I, which are bound weakly ($\sim$ -196 meV
in the +1 charge state for $\mu_{As}$, $\mu_{Mn}$=$\Delta$H(MnAs) and $\epsilon_F$=0 eV). We find antiferromagnetic 
spin arrangement in all $Q$=0, +1, +2 and +3 charge 
states considered. 
Thus, when these weakly bound S-I clusters are broken, depending on the charge state, there could be an
increase in the number of holes and consequently the ferromagnetic transition temperature.
On the other hand, S-I-S clusters appear to be stable and hence do not disintegrate under annealing.

\section{Summary} 

Under As-rich conditions, Mn prefers to substitute the Ga site. As the growth conditions become less
As-rich, or as extrinsic doping pushes $\epsilon_F$ towards and even below the VBM, the formation energy of
interstitial Mn becomes competitive with that of substitutional Mn. Under coherent epitaxial growth conditions, 
when MnAs precipitates are forced to be coherent with the zinc-blende
lattice, the formation energy of both substitutional and
interstitial decrease. At this point, the solubility is large enough to form clusers. We find that S-I-S clusters are
more stable than S-S-S clusters. S-I-S clusters are found to be strongly bound with respect to their 
constituents and exhibit partial or total hole compensation. While isolated 
Mn$_i$ behaves like a hole-killer and is expected to destroy ferromagnetism, 
in (Mn$_{Ga}$-Mn$_i$-Mn$_{Ga}$)$^0$, the Mn$_i$ is found to mediate
the ferromagnetic arrangement of spins on Mn$_{Ga}$.
The charged complex (Mn$_{Ga}$-Mn$_i$-Mn$_{Ga}$)$^{2+}$ has a similar ferromagnetic stabilization energy
on the two Mn$_{Ga}$ sites as in Mn$_{Ga}^0$-Mn$_{Ga}^0$ cluster without Mn$_i$ {\it almost as if Mn$_i$ did not exist}.
Thus ferromagnetism
in Mn doped GaAs arises from holes due to substitutional Mn$_{Ga}$, as well as 
from Mn$_{Ga}$-Mn$_i$-Mn$_{Ga}$ complexes.

This work was supported by the U.S. DOE, Office of Science, BES-DMS under contract no. DE-AC36-99-G010337.

\begin{figure}
\includegraphics[width=6.0in,angle=270]{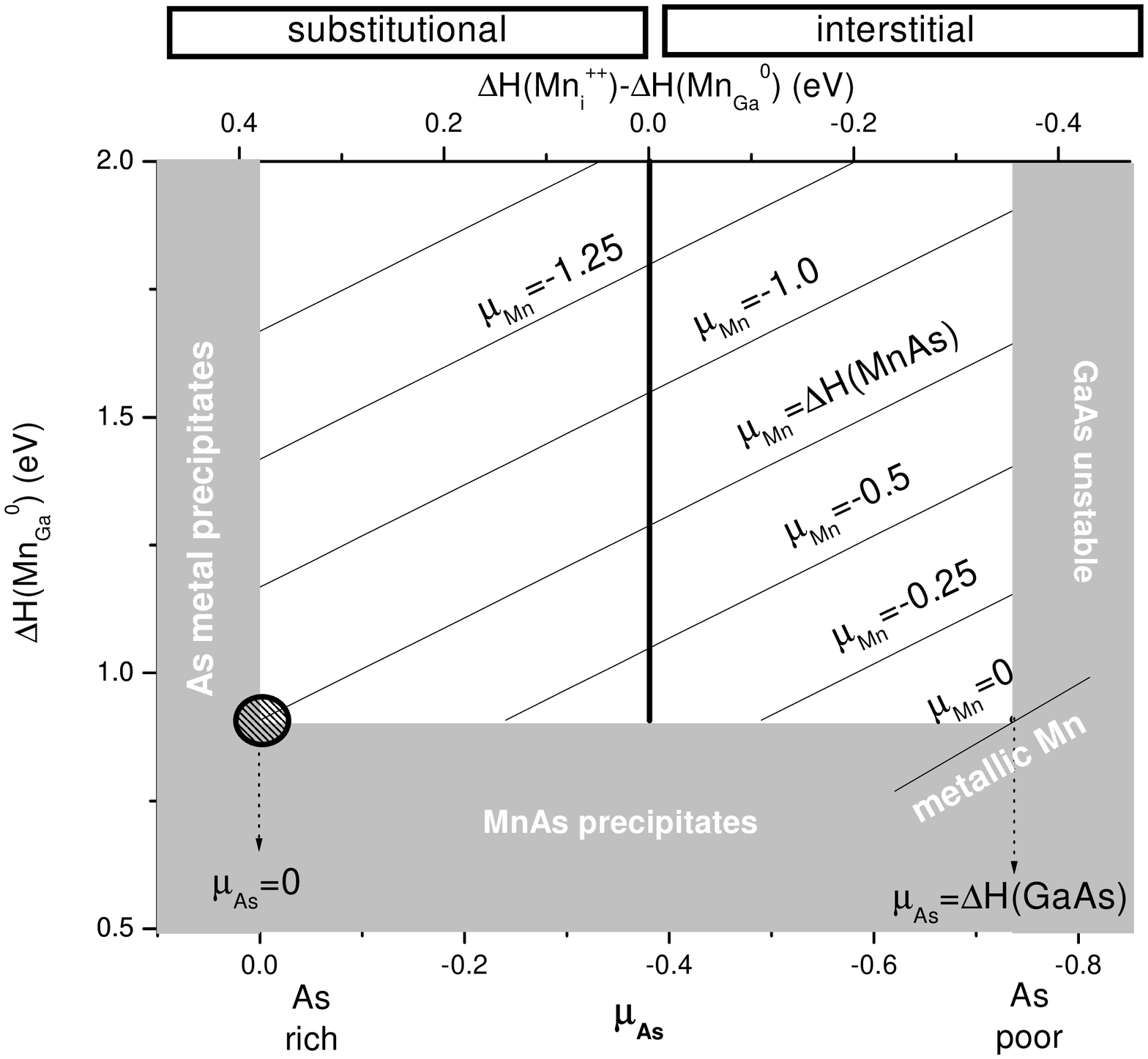}
\caption{The formation energy of Mn$_{Ga}^0$ (left y-axis) as well as the difference in 
formation energies of Mn$_i^{2+}$ and Mn$_{Ga}^0$ (top x-axis) 
are plotted as a function of $\mu_{As}$ (bottom x-axis)
for different values of $\mu_{Mn}$. Here $\epsilon_F$ is fixed at the the VBM of the host. 
Regions where there is precipitation of the elemental solids as well as MnAs are also shown.
}
\end{figure}

\begin{figure}
\includegraphics[width=6.0in,angle=270]{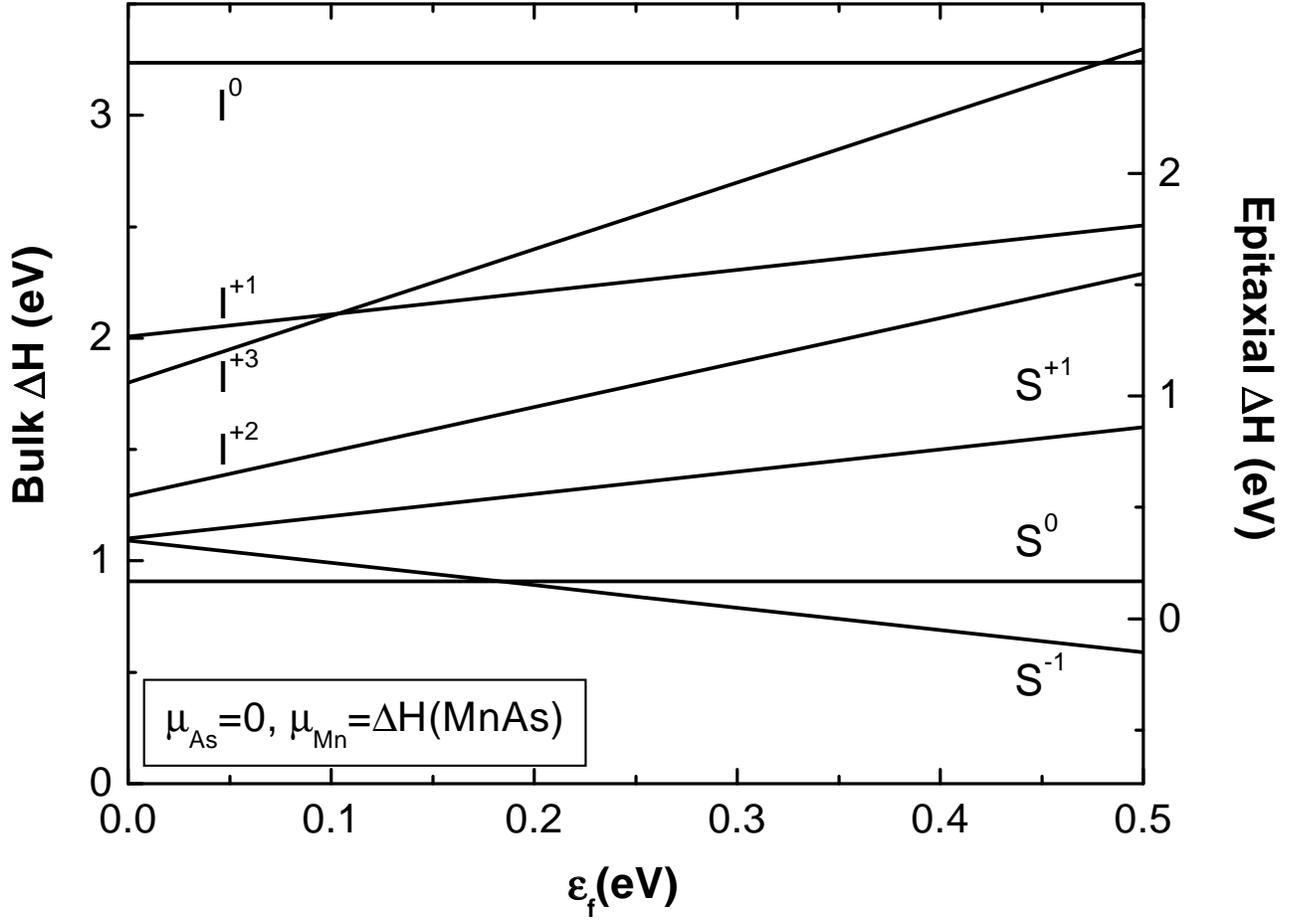}
\caption{ The bulk (left y-axis) as well as epitaxial (right y-axis) 
formation energies for different charge states of isolated substitutional (S) and isolated
interstitial (I) Mn calculated for a 64 atom supercell under As-rich conditions. 
Acceptor transition for 216 atom supercell (Table I) is E$_v$+0.13~eV. 
The chemical potentials are fixed at the points corresponding to the circle shown in Fig.~1.
}
\end{figure}

\begin{figure}
\includegraphics[width=6.0in,angle=270]{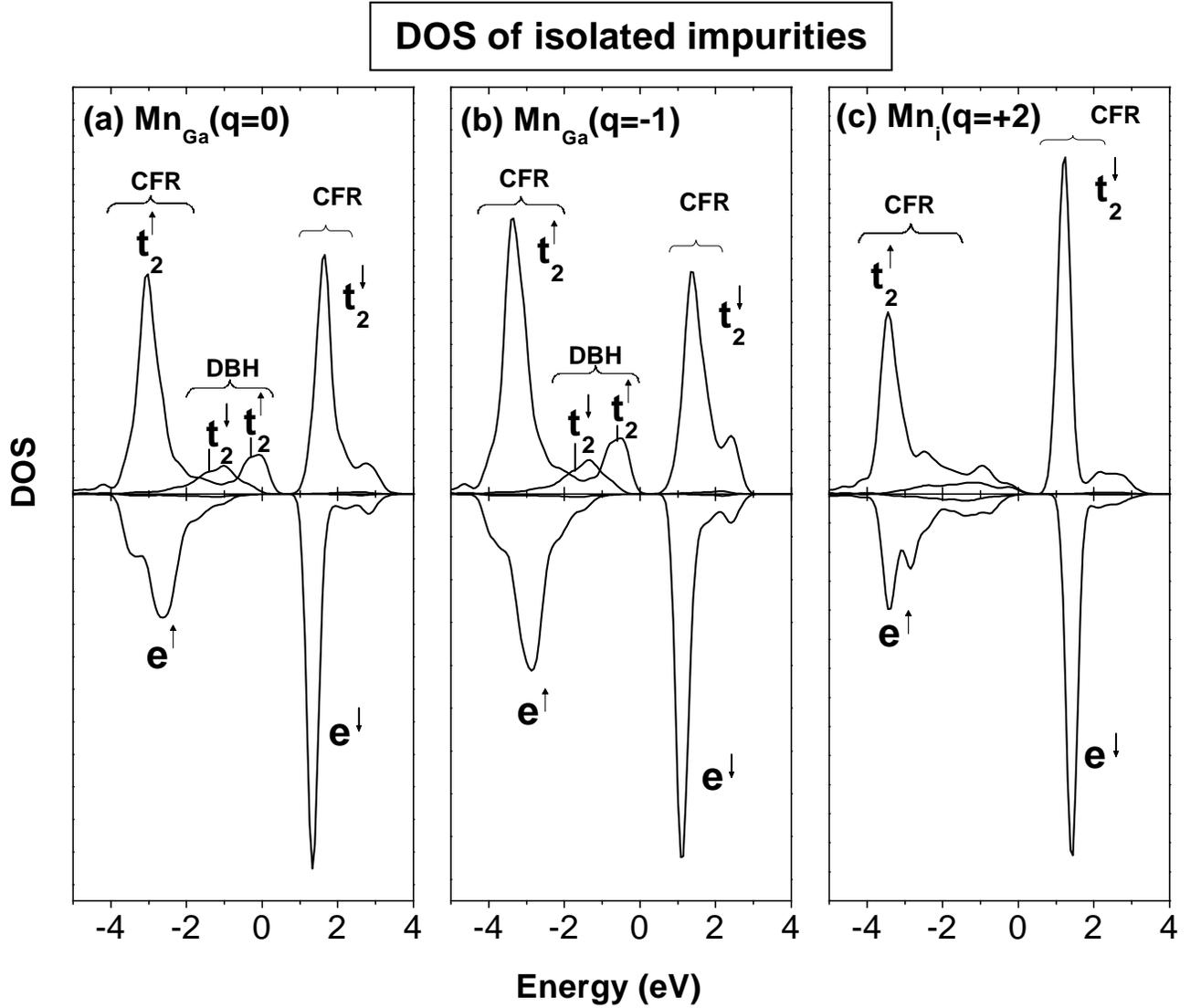}
\caption{The $t_2$ (upper panel) and $e$(lower panel) projected contributions to the Mn $d$ projected partial DOS
(a) for the $q$=0 and  (b) -1 states of Mn$_{Ga}$ as well as 
(c) the $q$=+2 charge state of Mn$_i$.}
\end{figure}

\begin{figure}
\includegraphics[width=5.5in,angle=0]{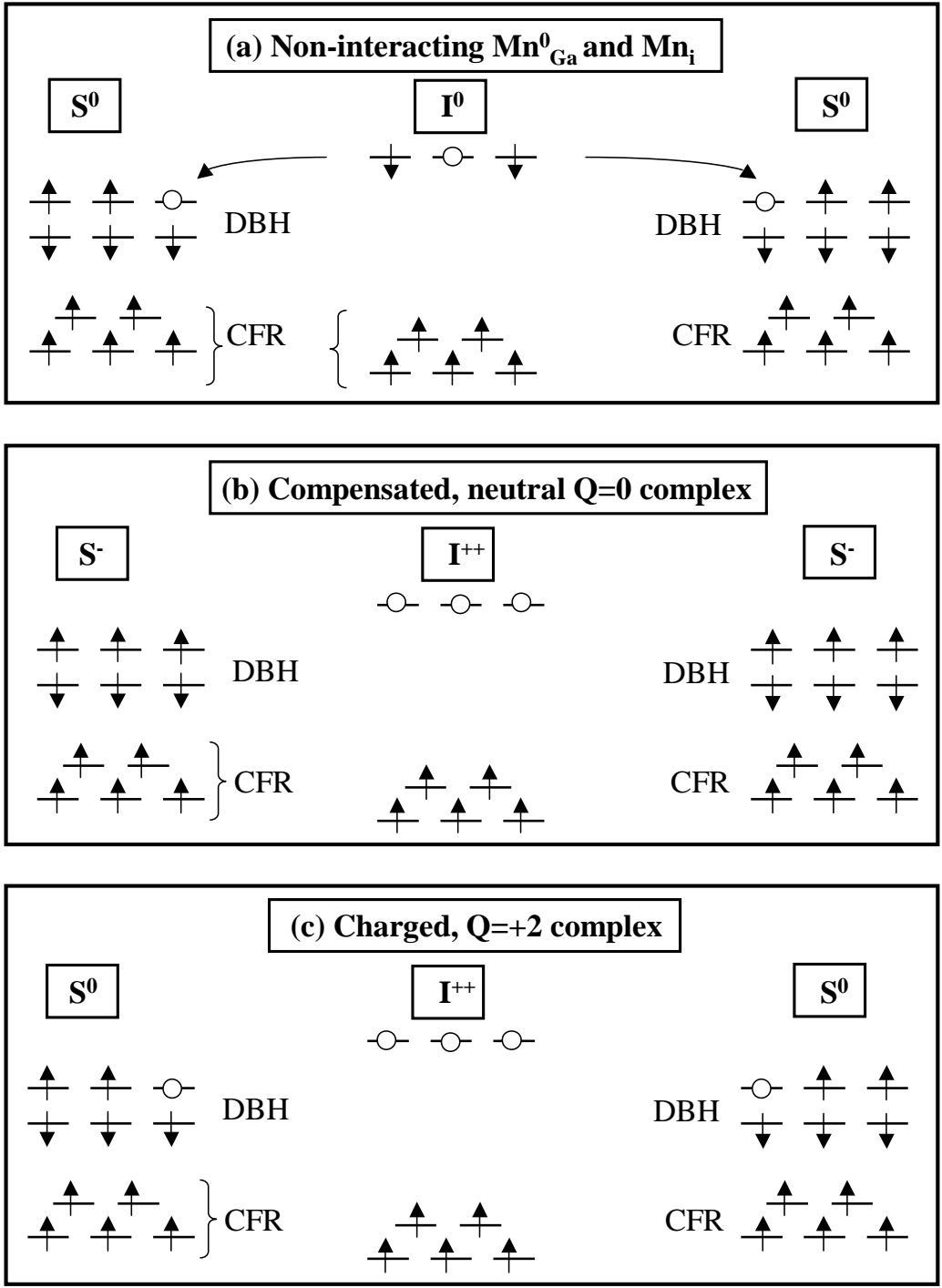}
\caption{ Schematic energy-level diagram for (a) neutral non-interacting substitutional (S) and interstitial (I) Mn impurities,
(b) the compensated S-I-S complex and (c) the doubly charged S-I-S complex involving 2 substitutional and one interstitial Mn, where Q is the total charge of the complex.
Open circles denote holes.
}
\end{figure}

\begin{figure}
\includegraphics[width=5.5in,angle=270]{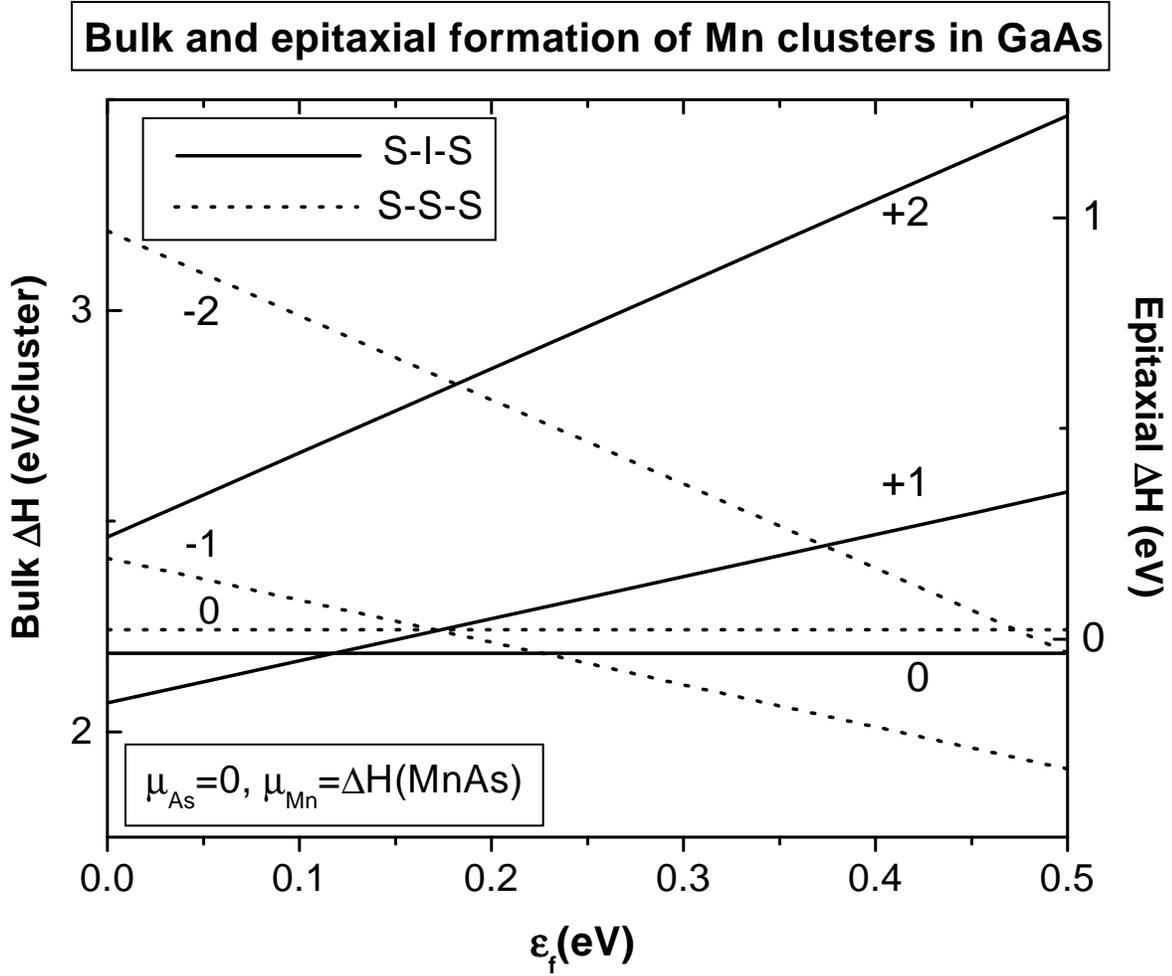}
\caption{ The bulk(left y-axis) as well as epitaxial (right y-axis) 
formation energies for different charge states of complexes involving
two substitutional and one interstitial Mn compared with three substitutional Mn 
calculated for a 64 atom supercell under As-rich conditions. 
The chemical potentials are fixed at the points corresponding to the circle shown in Fig.~1.
}
\end{figure}

\begin{figure}
\includegraphics[width=6.0in,angle=270]{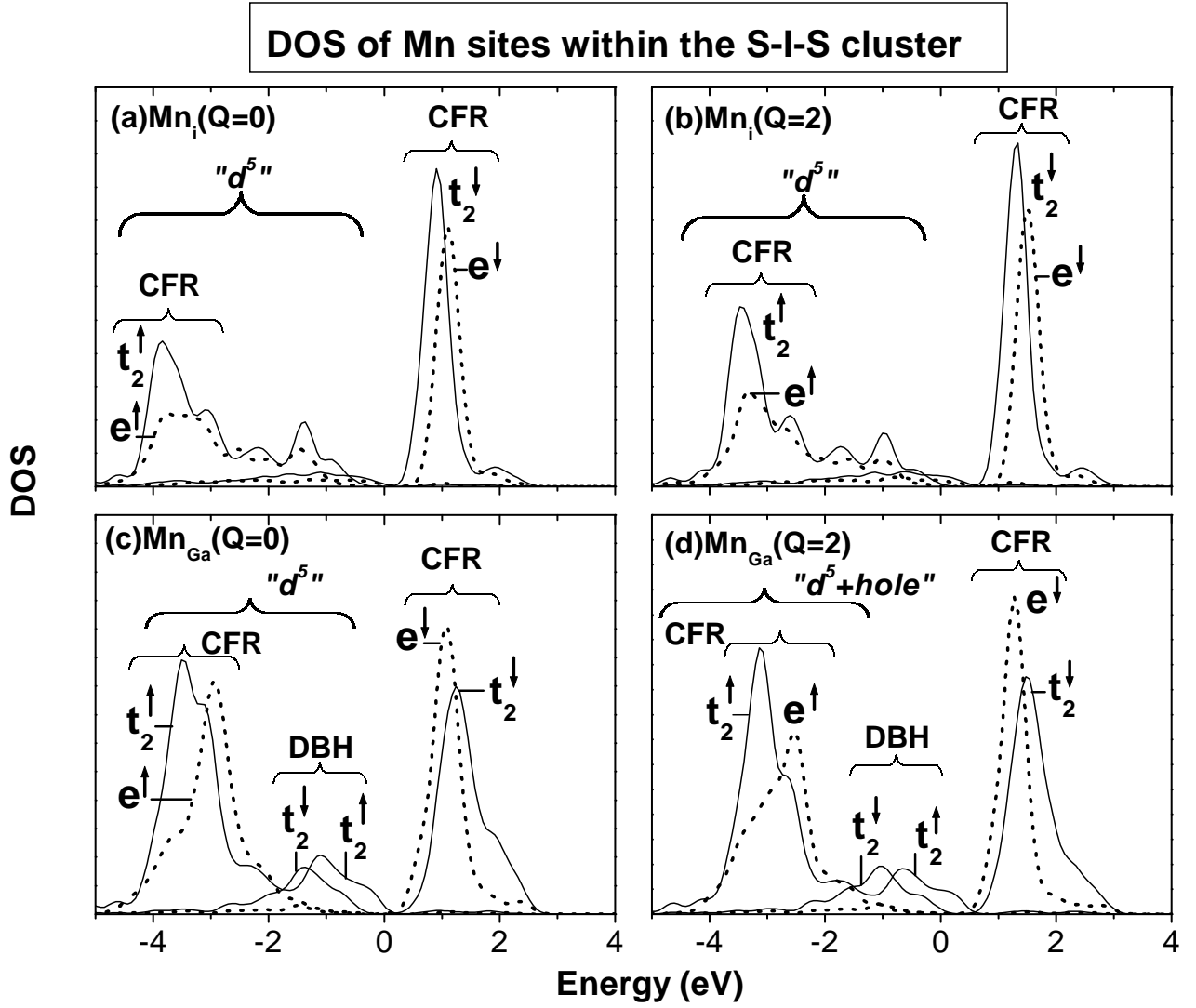}
\caption{The $t_2$ (solid line) and $e$(dotted line) projected contributions to the Mn $d$ projected 
partial density of states for Q=0 (left panels) and Q=+2 (right panels) of the complex projected 
onto Mn$_i$ (top panels) and Mn$_{Ga}$ (bottom panels).}
\end{figure}

\begin{figure}
\includegraphics[width=5.5in,angle=0]{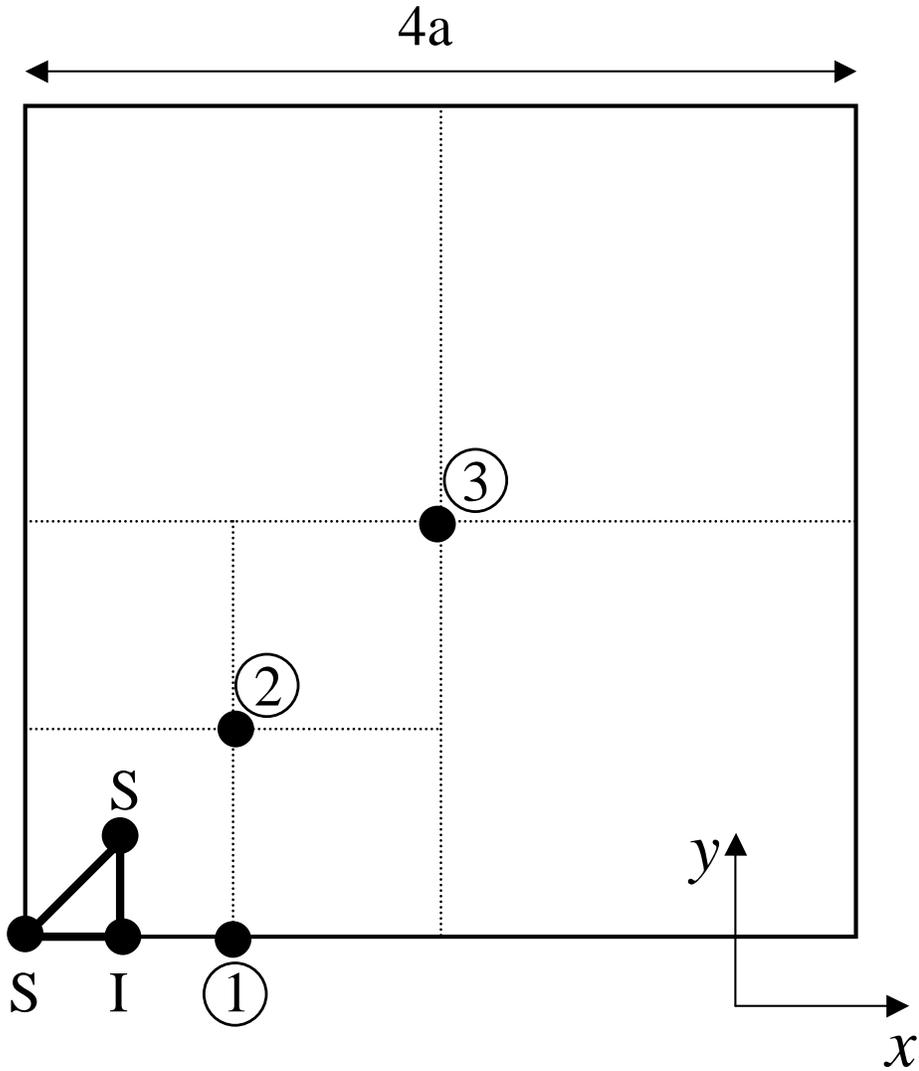}
\caption{A single face of the 256 atom supercell of GaAs used in our calculations, where $a$ is the cubic cell dimension.
Positions 1,2 and 3 considered for the isolated Mn$_{Ga}$ with respect to the cluster whose components
are labelled S and I.}
\end{figure}

\newpage
\begin{table}
\caption 
{ Acceptor transitions, Formation energies of Mn$_{Ga}$ and Mn$_i$ for 64 and 216 atom supercells of GaAs, with and without 
charge corrections } 
\begin{tabular}{l|l|l}
Quantity & 64 atom cell with (without) & 216 atom cell with (without) \\ 
         & charge correction (in eV) & charge correction  (in eV) \\ \hline
Mn$_{Ga}$(0/-) & 0.183  (0.094) & 0.133  (0.068) \\ \hline
$\Delta$H$_f$(Mn$_i^{2+}$)-$\Delta$H$_f$(Mn$_{Ga}^0$) & 0.382 (0.016) & 0.430 (0.17) \\ 
$\mu_{Mn}$=$\Delta$H(MnAs),$\mu_{As}$=0, $\epsilon_F$=0& ~ &  \\ \hline
$\Delta$H$_f$(Mn$_{Ga}^0$) & 0.908 +$\mu_{Ga}$ - $\mu_{Mn}$ & 1.261 +$\mu_{Ga}$ - $\mu_{Mn}$ \\ \hline
\end{tabular}
\end{table}

\begin{table}
\caption { The formation energy for different charge states of
isolated substitutional (Mn$_{Ga}$) as well as interstitial Mn
coordinated to four As atoms [Mn$_i$(As)] or to four Ga atoms
[Mn$_i$(Ga)] where $\mu_{\alpha}$ denotes the chemical potential
for atom $\alpha$.}
\begin{tabular}{c|c|c}
Charge state & \multicolumn{2}{c}{Formation energy} \\ \hline 
~  &   T$_d$ Mn$_i$(As) & T$_d$ Mn$_i$(Ga) \\ \hline
-1 &  3.81-$\mu_{Mn}$-$\epsilon_F$ & ~ \\
 0 & 2.45-$\mu_{Mn}$ & ~\\
+1 & 1.19-$\mu_{Mn}$+$\epsilon_F$ & 1.35-$\mu_{Mn}$+$\epsilon_F$ \\
+2 & 0.18-$\mu_{Mn}$+2$\epsilon_F$ & 0.49-$\mu_{Mn}$ +2$\epsilon_F$\\
+3 & 0.24-$\mu_{Mn}$+3$\epsilon_F$ & 0.55-$\mu_{Mn}$+$\epsilon_F$ \\ \hline
\end{tabular}
\end{table}

\begin{references}
\bibitem{compare} Our results for the 64 atom cell compare reasonably well with earlier
work of Erwin and Petukhov \cite{erwin} for substitutional and interstitial Mn in GaAs using 
54 atom cell. For example, the difference between 
the formation energies of Mn$_i^{2+}$ and Mn$_{Ga}^0$ with $\epsilon_F$=0 that we find is 0.38 
under As-rich conditions, while they 
obtain 0.44 for their 54 atom cell. Further they find Mn$_i$ related (+/2+) and (0/+) transitions at $\epsilon_F$ of
0.66 eV and 0.98 eV, while we get 0.72~eV and 1.23~eV respectively.
\bibitem{footnote1} 
The GGA optimised lattice constant for GaAs at the plane-wave basis cut-off of 13.3~Ry was 5.728 $\AA$,
to be compared with the 
experimental value of 5.65 $\AA$, while the value
using the LDA \cite{perdew_zunger} exchange functional is 5.601 $\AA$.
\bibitem{mn_reference}
Mn has a complex antiferromagnetic ground state which we have not considered in this work.
\end{references}
\end{document}